\documentclass[prl,a4paper,superscriptaddress,twocolumn,showpacs,amsmath,amssymb,floatfix,amstex]{revtex4}

\usepackage[dvips]{graphicx}
\usepackage{bm,pstricks,longtable}

\input{epsf}

\begin{document}

\title{Synchnonization, zero-resistance states and rotating Wigner crystal}


\author{A.D.Chepelianskii}
\affiliation{\mbox{Ecole Normale Sup\'erieure, 45, rue d'Ulm, 75231 Paris Cedex 05, France}}
\author{A.S.Pikovsky} 
\affiliation{\mbox{Department of Physics, University of Potsdam, 
  Am Neuen Palais 10, D-14469, Potsdam, Germany}}
\author{D.L.Shepelyansky}
\affiliation{\mbox{Laboratoire de Physique Th\'eorique, UMR 5152 du CNRS, 
Universit\'e  Toulouse III, 31062 Toulouse, France}}
\affiliation{\mbox{Department of Physics, University of Potsdam, Am Neuen Palais 10,  D-14469, Potsdam, Germany}}

\date{July 18, 2007}

\begin{abstract}
We show that rotational angles of electrons moving in two dimensions (2D) 
in a perpendicular  magnetic field can be synchronized
by an external microwave field which frequency 
is close to the Larmor frequency. The synchronization eliminates 
collisions between electrons and  thus creates 
a regime with zero diffusion corresponding to the zero-resistance states
observed in experiments with high mobility 2D electron gas (2DEG).
For long range Coulomb interactions electrons form a rotating
hexagonal Wigner crystal.
Possible relevance of this effect for planetary rings is discussed.
\end{abstract}


\maketitle
The discovery of microwave-induced resistance oscillations (MIRO)
\cite{zudov2001}
and of striking zero-resistance states (ZRS) 
of a 2DEG in a magnetic field 
\cite{mani2002,zudov2003}
attracted a great interest of the community.
A variety of theoretical explanations has been pushed forward
to explain the appearance of ZRS (see Refs. in \cite{isi}).
These approaches provide certain MIRO which at large
microwave power even produce a current inversion.
However, these theories do not give zero resistance, and
it is usually argued that ZRS are created as a result
of some additional instabilities 
which mysteriously compensate currents to zero.
Hence, a physical origin of ZRS still remains a puzzling problem.
 
In this work we suggest a generic physical mechanism
which leads to a suppression of electron-electron collisions and
creates ZRS. Its main element is the synchronization
phenomenon which has abundant manifestations in science,
nature, engineering and social life \cite{pikovsky,strogatz}. 
A simple picture of the effect is the following:
a microwave field excites electrons above the Fermi level and
switches on dissipation processes in energy, which compensate
microwave-induced
energy growth, thus creating a nonequilibrium steady-state distribution.
Due to this dissipation, when the microwave frequency $\omega$
is close to a resonance with the Larmor frequency $\omega_B$, 
the synchronization of the phases of Larmor
rotations of electrons with the phase of microwave field
is established.
In this way all electrons start to oscillate in phase - 
like male fireflies blink in phase on trees 
in Siam,  emitting rhythmic light pulses in order
to attract females \cite{pikovsky,strogatz}.
But compared to fireflies stationary sitting on trees,
the synchronization of moving electrons brings a new element 
not presented in the common 
synchronization science: due to synchrony the collisions
between electrons extinct what  leads to a drastic drop of
the collision-induced diffusion constant  $D$ and to creation of ZRS
(we note that $D$ is proportional to 
experimentally measured resistance  $R_{xx}$
since $R_{xy} \gg R_{xx}$ \cite{mani2002,zudov2003}).
A simple image of such synchronized electrons
is given by an ensemble of particles randomly distributed on
a 2D plane, which rotates as a whole on a Larmor circle of
radius $r_B=v_F/\omega_B$ with frequency $\omega_B$.
Indeed, in  such a rotating ensemble (plane) particles  never collide,
and we demonstrate below that this can happen with
2D electrons 
synchronized with a  microwave field phase in a magnetic field $B$.
The synchronization origin of ZRS allows one also to understand
qualitatively why ZRS exist only in high mobility samples.
Indeed, it is well known that synchronization remains
robust to a weak noise but 
disappears at strong one \cite{pikovsky},
hence a weak impurity scattering will not destroy ZRS.
It is also important to note that the above picture is based on classical
dynamics that has its grounds since in the experiments
\cite{mani2002,zudov2003} the Landau quantum level is large
$n_L \sim 100$. Thus, we start our analysis
with a classical mechanics treatment and 
will turn to a discussion of quantum effects later.

\begin{figure}[h]
\vglue -1.50cm
\epsfxsize=1.10\hsize
\epsffile{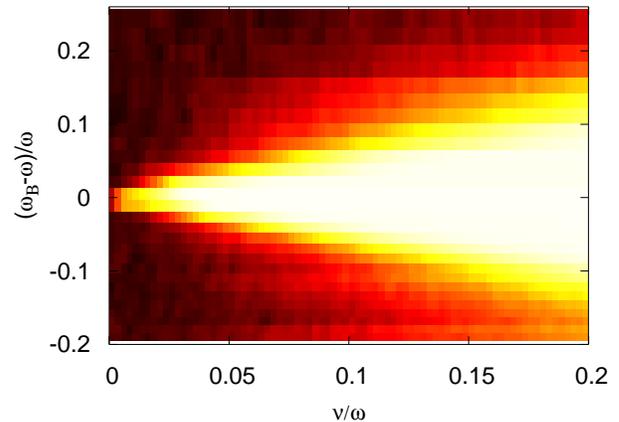}
\vglue -0.80cm
\caption{(color online)
Gray-scale  plot of the normalized diffusion rate
$D/D_0$ as a function of
frequency detuning $(\omega_B-\omega)/\omega$
and rescaled microwave field strength
$\nu/\omega$ with $\nu=f/m v_T$ where 
$v_T=\sqrt{2T/m}$ is the thermal velocity
and $D_0$ is the diffusion rate in absence
of microwave at $\omega_B=\omega$. 
The system parameters are:  $N=1000$
$N_c=4 \times 10^4$, $\omega \Delta t = 0.2$, $\omega \tau = 10$,
$\omega t = 500$, 
$L/r_B=10$, $D_0/D_c =0.12$ (with $D_c=v_T^2/\omega$, $\rho=N/L^2$ and 
$r_B$ taken at $\omega_B=\omega$, thus a number of particles inside
a Larmor circle is 
$N_B=\pi r_B^2 \rho= \pi \rho v_T^2/\omega^2 = 10\pi$,
$\omega = const$).
Color intensity is proportional to  $D/D_0$ 
(black for maximum $D/D_0 \approx 1.2$ and
white for minimum $D/D_0 = 0$).
\label{fig1}
}
\end{figure}

To justify the synchronization picture of ZRS described above we
perform extensive numerical simulations using two main models
of classical electrons (particles) with short range and Coulomb interactions.
In the simplest setup, we model particle dynamics with short range interactions 
in magnetic and microwave fields
with the Nos\`e-Hoover (NH) thermostat
(see e.g. \cite{hoover1,klages}) 
combined with interactions treated in the frame of
the mesoscopic multi-particle collision model (MMPCM) \cite{kapral2004}. 
The NH thermostat produces an effective friction $\gamma$
which keeps the average kinetic energy 
$\langle \mathbf{p}^2/2m \rangle$ equal to a given thermostat
temperature $T$ and 
equilibrates heating induced
by a microwave field $\mathbf{f}_{ac} = \mathbf{f} \cos{\omega t}$.
At the same time the MMPCM drives system toward  ergodic state with 
the equilibrium Maxwell distribution at a given
temperature. In this way the particle dynamics is described by the equations:
\begin{equation}
\label{eq1}
\mathbf{\dot q}_i = \mathbf{p}_i/m \; , \; 
\mathbf{\dot p}_i = \mathbf{F}_i+\mathbf{f}_{Li}+\mathbf{f}_{ac} - 
    \gamma \mathbf{p} \; , \;
\end{equation}
\vglue -0.80cm
\begin{equation}
\label{eq2}
\dot{\gamma} = [\langle \mathbf{p}^2 \rangle/(2mT) - 1]/\tau^2 
\end{equation}
where $\mathbf{q}_i, \mathbf{p}_i$ are the coordinate and 
the momentum of particle $i$,
$\mathbf{f}_{Li} = e \mathbf{[p}_i \mathbf{\times B]}/mc$ 
is the Lorentz force,
$\mathbf{F}_i$ is an effective force produced by particles collisions,
$\tau$ is the relaxation time in the NH thermostat
and $\langle \mathbf{p}^2 \rangle$ means average over all $N$ particles.
We usually consider the case of a
linearly polarized microwave field $\mathbf{f}_{ac}$
since numerical data give no significant dependence on polarization.
In numerical simulations $N$ particles 
are placed randomly on a square cell $L \times L$
which is periodically continued all over the plane. 
The collisions are treated in the 
MMPCM formalism, namely the main cell is divided into
$N_c$ small collision cells in which after a time step $\Delta t$ 
the velocities of particles are reshuffled randomly but keeping conserved
the momentum and energy of particles in the collision cell \cite{kapral2004}.
In absence of microwave radiation the system evolves to a usual thermal
equilibrium with the Maxwell distribution. The average
rate $D$ of particles diffusion in space is
computed via their displacements after a large time interval $t$.
In presence of the microwave field the diffusion rate $D$ is drastically changed
in the vicinity of the resonance $\omega_B \approx \omega$
as it is shown in Fig.~\ref{fig1} for typical values of 
parameters. 

Fig.~\ref{fig1} clearly shows the existence of a synchronization 
Arnold tongue inside which the diffusion drops to
zero (here as well as in numerical simulations below, 
its residual numerical value  
$D/D_0 \lesssim  10^{-8}$
is essentially determined
by roundoff errors and fit accuracy and is non distinguishable from zero). 
According to Fig.~\ref{fig1} the synchronization regime and the ZRS 
exist inside the detuning range
\begin{equation}
\label{eq3}
| \omega_B - \omega| \leq s f/(m v_T) \; ,
\end{equation}
where $v_T=\sqrt{2T/m}$ and a numerical constant $s \approx 0.7$. 
We note that $s$ is not sensitive to the relaxation time
$\tau$ which has been varied by an order of magnitude.
In fact the domain of ZRS  given by (\ref{eq3})
is very similar to a usual synchronization domain
for one particle \cite{pikovsky} which is also not
sensitive to the dissipation rate.
The origin of this similarity is rather clear:
the synchronization with the microwave field phase
eliminates collisions between particles, 
so that they move independently and hence the Arnold
tongue becomes the same as for one-particle synchronization.
The fact that in the ZRS the collisions are eliminated,
is confirmed by direct counting of the number of collisions
in the numerical code and by computation of the synchronization
parameter $S= \sum_{i<j} (\mathbf{v}_i - \mathbf{v}_j)^2/(N^2 v_T^2/2)$
which in the ZRS  drops down to $S \sim 10^{-10}$
being determined by roundoff errors. This means that 
all particles have the Larmor phase 
synchronized with the microwave field phase
while their positions in the coordinate space are disordered.
Outside of the ZRS particles continue to diffuse
with a rate  $D$ which is comparable with the unperturbed rate $D_0$.
At small values of $N_c$ and $\Delta t$ when the collision rate becomes 
rather large and $D_0 \sim D_c=v_T^2/\omega \approx v_T r_B$, 
the ZRS regime is destroyed.
Another model of collisions, in which 
the velocities of colliding particles are changed randomly
in a bounded relatively small scattering angle,
gives essentially the same result (\ref{eq3})
for the ZRS. 

To check the existence of the ZRS  in the case of long range
Coulomb interactions we use the molecular dynamics (MD) simulations
of a classical two-dimensional electron liquid as described in \cite{md}.
The results obtained in \cite{md} show that such an approach 
correctly describes plasmon modes in presence of a magnetic field even 
when the Coulomb energy $E_C = e^2/a$ is large compared
to classical temperature $T$. Here, $a=1/\sqrt{\pi \rho}$
is an average distance between electrons determined by the electron
density $\rho$. We ensured that our numerical code
with the Ewald resummation technique reproduces 
correctly the results presented in \cite{md}. 
To equilibrate the heating induced by
the microwave field we introduce in Eq.~(\ref{eq1})
an energy-dependent dissipation  
with $\gamma = \gamma_0 (E - E_F)/E_F$ for $E=p^2/2m > E_F$
and $\gamma = 0$ for $E<E_F$.  In such a way the dynamics
remains Hamiltonian for $E<E_F$ while above $E_F$ the
dissipative processes are switched on as it is usually the case
for 2DEG; thus $E_F$ plays a role of Fermi energy \cite{note}.
Usually we use $E_F/T \approx 2$ but the obtained results
are not sensitive to this ratio. The main part of simulations
is done at an intermediate interaction strength $r_s=E_C/E_F =0.3$
but we ensured that an increase(decrease) of $r_s$ by a factor
7(3) does not change qualitatively the results
(samples studied in \cite{mani2002,zudov2003} have $r_s \approx 2$). 
Also a variation of the dissipation rate $\gamma_0$
by an order of magnitude does not affect significantly the results
and we present  data at  $\gamma_0 \approx 0.7 v_F/a$.
The same is true for the total number of electrons
varied from 20 to 200 at $\rho=const$, thus we present data at $N=100$.

\begin{figure}[h]
\epsfxsize=0.82\hsize
\epsffile{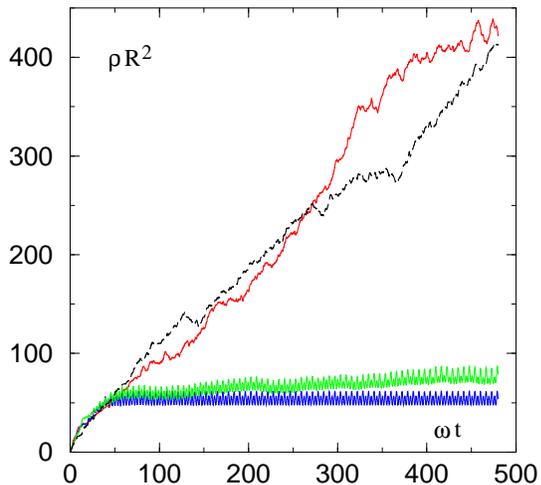}
\vglue -0.30cm
\caption{(color online) Dependence of electron square displacement $R^2$,
rescaled by electron density $\rho$,
on the rescaled time $\omega t$.
Here the Larmor frequency is $\omega_B=\omega$ at microwave
field strength $f=0$ (red top curve);
$f/(m v_F \omega) = 0.059$ 
($f a /E_F =0.02$) for $\omega_B=\omega$ (blue bottom curve),
$\omega_B=0.875\omega$ (second from top black dashed curve),
and $\omega_B=\omega$ with impurity scattering mean free path
$l_i  =  96 r_B$ (second from bottom green curve).
Total number of electrons is $N=100$ and 
$N_B = \pi \rho v_F^2/\omega^2 = 34.7$.
The linear fit gives the diffusion rates
$D/D_c = 0.089, 0.068, 0.0040, 9 \times 10^{-6}$
with $D_c=v_F^2/\omega$ 
(respectively for curves from top to bottom ordered at $\omega t =400$).
\label{fig2}
}
\end{figure}

A typical example of the dependence of average electron 
square displacement $R^2$
on time is shown in Fig.~\ref{fig2}. The introduction of microwave field
leads to the synchronization of electron Larmor phases
and to a drastic drop of diffusion rate at $\omega_B=\omega$, formally by
4 orders of magnitude; the synchronization parameter $S$
drops down to $S \approx 10^{-11}$ in this case
that means that collisions are completely switched off.
A shift in the Larmor frequency $\omega_B=0.875\omega$
destroys synchronization and diffusion $D$ is restored being
close to its unperturbed value $D_0$ at $f=0$. An introduction of
a weak noise linked to impurity scattering with a scattering time
$t_i$ and mean free path $l_i=v_F \tau_i$
leads to a finite diffusion rate $D$ which
is however much smaller than $D_0$ until
$l_i \gg r_B$ (see Fig.~\ref{fig2}). A decrease of the mean free path
down to $l_i \approx 10r_B$ destroys synchronization and 
restores a diffusion with rate $D \approx D_0$.
We note that $l_i \approx 100 r_B$ approximately corresponds
to experimental conditions in \cite{mani2002,zudov2003}.
\begin{figure}[h]
\epsfxsize=0.80\hsize
\epsffile{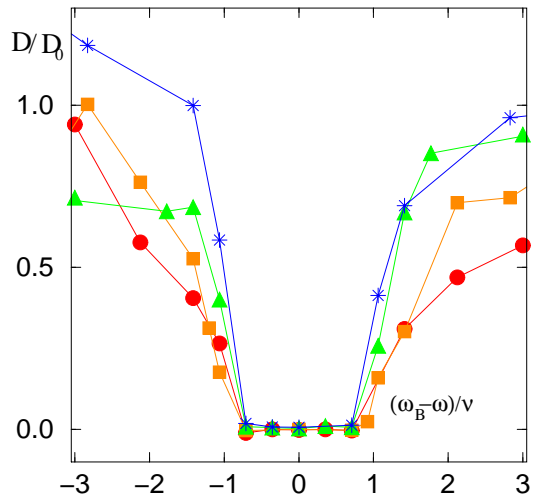}
\vglue -0.30cm
\caption{(color online) 
Dependence of rescaled diffusion rate
$D/D_0$ on the rescaled frequency difference
$(\omega_B - \omega)/\nu$.
Here $\nu=f/mv_F$, $D_0$ is diffusion rate in absence of microwave
at $\omega_B=\omega$,
$fa /E_F =0.02$ and number of electrons in a Larmor circle is
$N_B = $ 2 (stars), 8 (triangles), 34.7 (squares), 138.8 (points)
with $D_0/D_c =$ 0.054, 0.089, 0.12,  0.14 and
$D_0/v_F a = $ 0.20, 0.35, 0.53, 0.64 respectively.
Total number of electrons is $N=100$, $L=\sqrt{N/\rho} \approx 17.72 a$.
\label{fig3}
}
\end{figure}

The dependence of $D$ on the frequency detuning is 
shown in Fig.~\ref{fig3}. The numerical data for Coulomb interactions
between electrons show that the ZRS  exist inside the 
synchronization window near the resonance $\omega_B \approx \omega$
with the width given by Eq.~(\ref{eq3}) where $v_T$ should be replaced
by $v_F$ and $s \approx 0.8$. 
Inside the ZRS  the diffusion drops practically
to zero as discussed above.
The validity of the relation
(\ref{eq3}) shows that the effect is not very sensitive to
the type of interactions between particles.

\begin{figure}[h]
\vglue -0.30cm
\epsfxsize=1.0\hsize
\epsffile{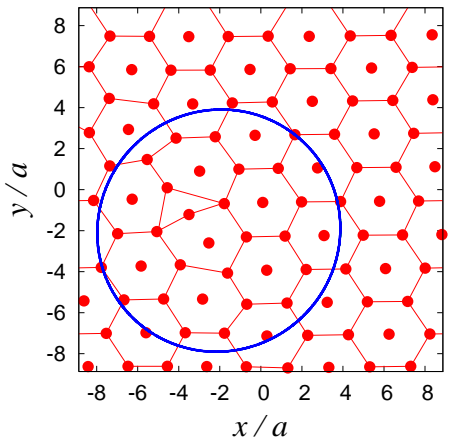}
\vglue -0.30cm
\caption{(color online)
Instant image of the rotating Wigner crystal  formed by
$N=100$ electrons (points) in a periodic cell
with $L=\sqrt{N/\rho} \approx 17.72 a$,
$\omega t = 480$, $\omega_B=\omega$, $f a/E_F=0.02$ and $N_B=34.7$
(as in Fig.~\ref{fig2}, bottom curve); the circle shows an orbit
of one electron for $240 \leq \omega t \leq 480$;
lines are drawn to adapt an eye
showing a hexagonal crystal with a defect.
\label{fig4}
}
\end{figure}

However, the long range nature of Coulomb interactions significantly modifies 
the structure of the ZRS configuration: for short range interactions
particles are distributed over the plane in a disordered way,
while for the Coulomb interactions electrons
form a hexagonal Wigner crystal as it is shown in Fig.~\ref{fig4}.
The whole crystal (as well as each electron) is rotating
in the plane with the frequency $\omega \approx \omega_B$
and rotation radius $r_B=v_F/\omega_B$. 
A remarkable property of the rotating Wigner crystal is that
formally it is formed at a rather small parameter $r_s \approx 0.3$
while the usual Wigner crystal requires 
$r_s$ values by more than two orders of magnitude larger
\cite{ceperley}. We attribute this to synchronization
of electron Larmor phases with the microwave field phase,
what eliminates  collisions between
electrons and suppresses fluctuations, thus yielding 
an effectively large  $r_s$ in the rotating frame. 
In the crystal all Coulomb forces acting
on  an electron are compensated, thus 
the size of synchronization domain  in frequency range
given by Eq.~(\ref{eq3}) is essentially the same as for
one-particle synchronization and 
is practically independent of dissipation rate $\gamma_0$
\cite{pikovsky}.

In conclusion, we have suggested a generic 
mechanism which  for 2D particle rotational dynamics
(e.g the Larmor rotation, but it may be also relevant 
for other 2D systems like planetary rings) 
produces synchronization of
rotational angles of all particles with the phase
of external driving periodic field. 
As a result 
a rotating Wigner crystal is created 
and a collisional diffusion is suppressed by several orders of magnitude.
The collective crystal structure also suppresses the diffusion
due to impurities.
We propose  that this effect explains the appearance of ZRS in 
2DEG observed  in \cite{mani2002,zudov2003}. 
According to Eq.~(\ref{eq3}) the relative size
of ZRS plateau is 
$\Delta \omega/\omega \approx 2\nu/\omega \approx f v_F /\omega E_F$
that for experiments \cite{mani2002,zudov2003}
with $E_F \sim 100 K^{o}$, $v_F \sim 3 \times 10^7 cm/s$
and $\omega/2\pi = 35 GHz$  gives $\Delta \omega / \omega \approx 0.1$
if the field strength  acting on an electron is
$f/e \approx 5 V/cm$.
This relative width is in a reasonable agreement with the
experimental results \cite{mani2002,zudov2003,bykov1}
where  unfortunately an exact value of $f$ is not known.
The synchronization energy scale
$E_S \sim f r_B \sim 10 K$
and the crystal Coulomb energy $E_C \sim 200 K$
might be the origin of large energy
scale $E_A \sim 10 K$ 
in the ZRS activated transport \cite{mani2002,zudov2003}.
An important discrepancy from the experiments is that
our theory gives synchronization only near the main resonance
$\omega_B/\omega \approx 1$ while in the experiments
ZRS  exist also near integer low  resonances $j$ with
$\omega_B/\omega \approx 1/j$. We suppose that these resonances
may appear due to an additional effective 2DEG potential
and surface  modulation in space. 
This may generate higher harmonics of the Larmor
motion and produce synchronization also at integers $j > 1$.
Such a modulation apparently appears during molecular epitaxial growth
\cite{bykov2}. This modulation also produces a frequency shift
in the rotational frequency that may be responsible for
a resonance shift of the ZRS domain compared to the
Larmor resonance (see \cite{mani2002,zudov2003,bykov1}).
A coherent rotation of electrons in the crystal generates
a rotating magnetic field
$B_W \sim \mu_0 e v_F \rho \sim 1 G$
parallel to 2DEG which can be detected experimentally.

Our theory is based on the classical dynamics and it is crucial 
to analyze the contribution of quantum effects. 
In principle it is known that at small effective values of
Planck's constant $\hbar_{eff}$ the synchronization is preserved
while at large values $\hbar_{eff}$ it is destroyed by
quantum fluctuations \cite{zhirov}. For 2DEG $\hbar_{eff} \sim 1/n_L$
and at $n_L \sim 100$ it is natural to expect that
the synchronization is robust against quantum fluctuations.
However, a reduction of $n_L$ by an order of magnitude
due to an increase of $\omega$ to a THz range may 
significantly enhance quantum noise and destroy ZRS.
Further theory development is required to study quantum effects
properly. The most important question is about the amount
of electrons which are involved  in the rotating Wigner crystal.
Indeed, our classical studies show that all electrons
are involved in this state but in the quantum case
it is rather possible that only a finite fraction of
electrons near the Fermi level contributes to the
rotating crystal, while all other electrons will stay
as a non-interacting background.

Finally we make a conjecture that the mechanism described here may
be responsible for enormously long life time ( $\sim 10^{12}$ rotations) 
and sharp edges of planetary rings (e.g. $\sim 10 m$ for Saturn)
\cite{fridman,spahn}.
Indeed, a temperature there is very low and
in the rotational frame the 2D  dynamics of particles 
is similar to  motion of electrons in a magnetic field \cite{fridman}. Hence,
moons inside a ring and near to a resonance may produce
synchronization and diffusion suppression with emergence of ZRS in space.

We thank A.Bykov, J.Schmidt and F.Spahn for useful discussions.
This work was supported in part by the ANR PNANO project $\;\;\;$ MICONANO.

\vglue -0.30cm


\end{document}